\documentclass[prl,twocolumn,amsmath,amssymb,floatfix,footinbib,bibnotes,longbibliography]{revtex4-1}

\pdfoutput=1

\usepackage[colorlinks=true,citecolor=blue,linkcolor=magenta,hypertexnames=false]{hyperref}
\usepackage{amsmath}
\usepackage{graphicx}
\usepackage{dsfont}
\usepackage{changepage}
\usepackage{fancyhdr}
\usepackage{amsthm, amssymb}
\usepackage{float}
\usepackage{array}
\usepackage[usenames,dvipsnames]{color}
\usepackage{slashed}

\newcommand{\bj}{{\bf j}}
\newcommand{\bk}{{\bf k}}
\newcommand{\bq}{{\bf q}}

\newcommand{\bK}{{\bf K}}
\newcommand{\bE}{{\bf E}}
\newcommand{\bA}{{\bf A}}

\newcommand{\bp}{{\bf p}}

\newcommand{\ve}{{\varepsilon}}

\def \tr{{\mbox{tr~}}}

\def \be{\begin{equation}}
\def \ee{\end{equation}}
\def \bea{\begin{eqnarray}}
\def \eea{\end{eqnarray}}

\begin{document}

\title{Interactions remove the quantization of the chiral photocurrent at Weyl points}

\author{Alexander Avdoshkin$^{1}$}
\thanks{These two authors contributed equally.}
\author{Vladyslav Kozii$^{1,2}$}
\thanks{These two authors contributed equally.}
\author{Joel E. Moore$^{1,2}$}

\affiliation{ {
1. Department of Physics, University of California, Berkeley, CA 94720, USA \\
2. Materials Sciences Division, Lawrence Berkeley National Laboratory, Berkeley, CA 94720, USA  }}

\begin{abstract}
{

The chiral photocurrent or circular photogalvanic effect (CPGE) is a photocurrent that depends on the sense of circular polarization.  In a disorder-free, noninteracting chiral Weyl semimetal, the magnitude of the effect is approximately quantized with a material-independent quantum $e^3/h^2$ for reasons of band topology.  We study the first-order corrections due to the Coulomb and Hubbard interactions in a continuum model of a Weyl semimetal in which known corrections from other bands are absent.  We find that the inclusion of interactions generically breaks the quantization.  The corrections are similar but larger in magnitude than previously studied interaction corrections to the (nontopological) linear optical conductivity of graphene, and have a potentially observable frequency dependence.  We conclude that, unlike the quantum Hall effect in gapped phases or the chiral anomaly in field theories, the quantization of the CPGE in Weyl semimetals is not protected but has perturbative corrections in interaction strength.

 }
\end{abstract}
\maketitle

The quantization of physical observables has become a cornerstone in condensed matter physics for the past few decades, guiding theoretical and experimental efforts across a wide range of fields. Starting from the discovery of quantum Hall effect, it led to multiple breakthroughs in our understanding of quantum systems. For example, identifying the quantization of Berry phase led to the discovery of topological insulators~\cite{HasanKane2010}.  A few examples of quantization of electronic and optical properties have been identified in metallic systems as well, such as the universal optical conductivity and optical transmittance in graphene~\cite{GusyninSharapovCarbotte2006,Falkovsky2007,Kuzmenkoetal2008,Nairetal2008} and, more recently, the circular photogalvanic effect (CPGE) in Weyl semimetals~\cite{deJuanGrushinMorimotoMoore2017} and crystals with multifold nodal  fermions~\cite{Ezawa2017,HuangMultiWeyl2017,Lepori2018,Flickeretal2018}.

One of the crucial questions in the study of topological materials, from both experimental and theoretical perspectives, is whether the quantized features are robust against interactions and disorder. In most cases, weak interactions do not destroy or qualitatively change gapped topological phases~\cite{Rachel_2018}. In particular, the quantum Hall and quantum anomalous Hall conductivities are known to preserve the quantized value even in the presence of a weak interaction~\cite{AvronSeiler1985,niuthoulesswu1985,Ishikawa1986,Hastings2015,COLEMAN1985184,HaldaneHubbard2016,Giuliani2017}, which is intimately related to the topological nature of the  effect~\cite{AvronSeiler1985,Ishikawa1986,Hastings2015} and ultimately its connection to adiabatic transport~\cite{laughlin1981}.

For Weyl fermions, the effect of interactions was exhaustively studied in the context of the chiral anomaly --- nonconservation of the chiral charge without an explicit breaking of the chiral symmetry~\cite{Adler1969,BellJackiw1969}. It has long been known that the anomaly is not renormalized by interactions that  do not explicitly break the chiral symmetry~\cite{AdlerBardeen1969,Giulianetal2019}. Analogously to the quantum Hall effect, this nonrenormalizability is deeply rooted in the topological nature of the chiral anomaly~\cite{ZyuzinBurkov2012,Burkov_2015,Burkov2018}.  The chiral anomaly, however, leads to quantization of the chiral current, which in condensed matter is not the observable electrical current but rather a pumping between Weyl nodes, and hence it has not yet been possible to observe the quantization despite various proposals~\cite{parameswaran2014,
kolodrubetz2016}.

In this work, we study the effect of electron-electron interactions on the CPGE, another quantized response in nodal semimetals~\cite{deJuanGrushinMorimotoMoore2017}. The CPGE is the production of a dc current by a circularly polarized light incident on a surface of the material~\cite{AversaSipe1995,SipeShkrebtii2000,NastosSipe2010,Spivaketal2009,Moore2010,levchenko2017,GolubIvchenko2018}. In particular, the CPGE is the part of photocurrent that switches sign depending on the sign of the light polarization. This is a nonlinear response, second order in electric field, and hence requires the breaking of inversion symmetry; such responses can have topological content~\cite{Morimoto2016}.  In Weyl semimetals that are also free of mirror symmetries,  the CPGE becomes approximately quantized over some range of frequencies.  As found in Ref.~\cite{deJuanGrushinMorimotoMoore2017}, the intrinsic contribution from a single Weyl point to the CPGE, an injection current $\bj$, is quantized and has the value 
\be
\frac{d \bj}{dt} = \beta_0(\omega) \,  \bE_{\omega} \times \bE_{-\omega}, \qquad \beta_0(\omega) = i \, \frac{\pi e^3}{3 h^2}C, \label{Eq:CPGEquant}
\ee
where $e$ is the electron charge, $h = 2\pi \hbar$ is the Planck's constant, and $C$ is the topological charge of the node.  The important prerequisite for this result is the absence of inversion and mirror symmetries.  Then nodes of different chiralities are located at different energies. Consequently, for a certain frequency range, one node contributes exactly the quantized value~(\ref{Eq:CPGEquant}) from the transitions across the Weyl point, while the second one does not have such interband transitions because of Pauli blocking. The intraband contribution to the CPGE current originating, e.g., from the indirect disorder-assisted transitions and allowed even at low frequencies, is typically at least an order of magnitude smaller and governed by previous semiclassical calculations~\cite{Spivaketal2009,Moore2010,levchenko2017,GolubIvchenko2018}. 
So while the CPGE involves generation of a three-dimensional current density from two powers of electromagnetic field, like the chiral anomaly, unlike the chiral anomaly it can be observed in the overall electrical current, not the chiral current between nodes.  Remarkably, this effect was recently predicted~\cite{RhSiprediction2017}, and the distinctive frequency dependence observed~\cite{RhSiexperiment2019}, in the chiral Weyl semimetal RhSi.

We show using a minimal continuum model of a chiral Weyl semimetal that generic interactions give corrections to the perfect quantization of the CPGE, in contrast to the chiral anomaly. This model has the feature that corrections from other pieces of the Fermi surface are absent and the CPGE is exactly quantized without interactions for a range of frequencies. Furthermore, as was shown in Ref.~\cite{deJuanGrushinMorimotoMoore2017}, the quantization within the noninteracting two-band model is quite robust in the sense that it does not depend on such material/model-specific details as the Fermi velocity, tilt of the node, or the exact position of the chemical potential, and is given by Eq.~(\ref{Eq:CPGEquant}). While the topological charge of the nodes $C$, when properly defined, is not affected by weak interactions~\cite{Volovik1988,Gurarie2011}, we find that the universal proportionality between the CPGE coefficient $\beta$ and the topological charge, Eq.~(\ref{Eq:CPGEquant}), does not hold in the presence of interactions. Using the low-energy field theory suitable for Weyl fermions, we demonstrate that the CPGE response acquires a nonuniversal correction even at weak coupling, in the sense that this correction depends on such material-specific parameters as the Fermi velocity or dielectric constant~\footnote{This is in contrast to the Coulomb interaction correction to the optical conductivity in Weyl semimetals, which was found to be universal~\cite{RoyJuricic2017}.}.  So while $e^3/h^2$ remains the natural scale for the CPGE response, there are potentially observable interaction effects that need not be small in real materials.  We use the Hubbard and screened Coulomb potentials as examples.

Our results imply that the CPGE is an example of a quantized response which is not protected by topology beyond the noninteracting limit, and hence gets renormalized by arbitrarily  weak interactions. In some sense, this scenario is similar to the effect of the interaction corrections to the (nontopological) optical conductivity in graphene.  While the noninteracting consideration leads to the quantized value $e^2/4\hbar$~\cite{GusyninSharapovCarbotte2006,Falkovsky2007,Kuzmenkoetal2008,Nairetal2008}, the presence of interactions is known to contribute additional nonuniversal  correction~\cite{HerbutVafek2008,VafekHerbut2010,Mishchenko_2008,SheehySchmalian2009,Schmalian2016,SodemannFogler,ResensteinLewkowiczManiv2013,Teber2017,TeberKotikov2018}. Similar results have been recently obtained for the optical conductivity in nodal-line semimetals~\cite{nodal-lineoptcond2019}.

The calculation of  the numerical coefficient for the interaction correction in graphene turned out to be a nontrivial task. Originally, three different values of this coefficient were obtained for the hard-cutoff, soft-cutoff, and dimensional regularization schemes~\cite{HerbutVafek2008,VafekHerbut2010,Mishchenko_2008,SheehySchmalian2009,Schmalian2016},
leading to an intensive discussion regarding the choice of the correct one. The reason for such a peculiar behavior is rooted in the ultraviolet anomaly~\cite{Mishchenko_2008}: when applied na\"{i}vely, different approaches differently account for the high-energy states, resulting in different answers. It was shown later that, when the renormalization procedure is performed carefully, the soft-cutoff and dimensional regularizations lead to the same answer~\cite{Teber2017,TeberKotikov2018}. We also encounter the same anomaly in our study. We find that the results obtained within the soft-cutoff and the dimensional regularization procedures agree with each other, while the scheme with the hard cutoff, which implies neglecting the electron states with momenta exceeding certain ultraviolet value, leads to a different answer. This is somewhat natural, since the presence of a hard cutoff violates the Ward-Takahashi identity and incorrectly accounts for the contribution from the high-energy states, leading to a result which is only qualitatively correct.

{\it Quantization of the CPGE in the absence of interaction. --- } Before presenting the main results of our paper, we first reproduce the result for the noninteracting problem~\cite{deJuanGrushinMorimotoMoore2017} using the framework of Feynman diagrams. The detailed derivation of the second-order response within the Keldysh formalism is given in Ref.~\cite{JoaoLopes2018} (see also Refs.~\cite{RammerSmith1986,ParkeretalDiagrams2019}). In this work, however, we find it more convenient to use the Matsubara imaginary time formalism~\cite{RostamiJuricic2019}, which is equivalent to the Keldysh approach.

\begin{figure}
  \centering
  \caption{ Schematic picture of two Weyl nodes of opposite chirality separated by energy $|\mu_1| + |\mu_2|$. The quantization of the circular photogalvanic effect in the noninteracting material occurs provided $2|\mu_1| < \omega < 2|\mu_2|$. }
  \includegraphics[width=1.\columnwidth]{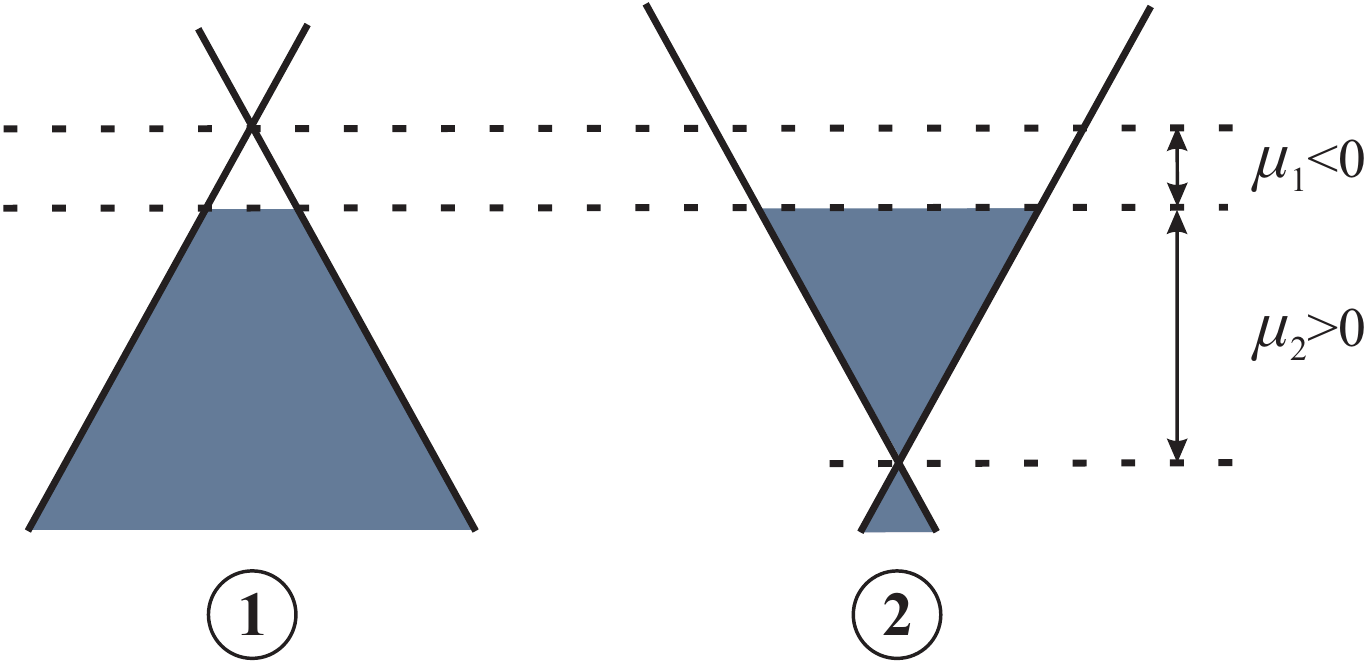}
  \label{Fig:twonodes}
\end{figure}

We start with a noninteracting system of two identical Weyl nodes of opposite chirality separated by energy $|\mu_1| + |\mu_2|$, as shown in Fig.~\ref{Fig:twonodes}. We assume for definiteness that the chemical potential for the first node is negative, $\mu_1 < 0$, while for the second node it is positive, $\mu_2 >0$. The low-energy Hamiltonian of the system then takes the form 

\begin{align}
H_0 = \sum_{\bk} &\psi_{1 \bk}^\dagger \left( v_F \bk \cdot \boldsymbol{\sigma} - \mu_1 \right) \psi_{1 \bk} + \nonumber \\ + &\psi_{2 \bk}^\dagger \left( - v_F \bk  \cdot \boldsymbol{\sigma} -  \mu_2 \right) \psi_{2 \bk}, \label{Eq:H0}
\end{align}
where $\psi_1$ and $\psi_2$ are two-component fermion spinors describing the states near the first and second node, respectively, ${\boldsymbol \sigma}$ is a vector of pseudospin Pauli matrices, and $v_F$ is the Fermi velocity. Here and in what follows, we set $\hbar=1$ for brevity, unless explicitly stated otherwise. The different sign of the Fermi velocities reflects the fact that the nodes have different chiralities.

We assume that the nodes are well separated in  momentum space, and consequently the contribution to the (uniform) photocurrent can be calculated separately for each node. For definiteness, we focus on the first node for now.  The expression for the second-order photocurrent reads as

\be
j^\gamma(\Omega) = \frac{\chi^{\alpha \beta \gamma}(\omega_1, \omega_2) +  \chi^{\beta \alpha  \gamma}(\omega_2, \omega_1) }{\omega_1 \omega_2}  E^{\alpha}_{\omega_1} E^{\beta}_{\omega_2}, \label{Eq:currentgeneral}
\ee
where $\Omega \equiv \omega_1 + \omega_2$, and the factors $\omega_{1,2}$ in the denominator originate from the relation between the electric field and the vector potential $\bE_\omega = i \omega \bA_\omega$. The analytical expressions for the tensor $\chi(i\omega_1, i\omega_2)$  in Matsubara frequencies is given by~\cite{JoaoLopes2018}

\begin{align}
&\chi^{\alpha \beta \gamma}(i \omega_1, i\omega_2) = T\sum_{\ve_n} \int \frac{d^3 k}{(2\pi)^3} \text{tr} \,\left[ \hat  j^{\alpha} \, G(i\ve_n - i\omega_1, \bk) \times \right. \nonumber \\  & \left.  \times \hat j^{\beta}\, G(i\ve_n - i  \Omega, \bk) \, \hat j^{\gamma} \, G(i\ve_n, \bk)\right], \label{Eq:chi}
\end{align}
with $\ve_n = \pi T (2n + 1)$ and $T$ is temperature. The current operator in this expression equals 

\be
\hat j^{\alpha} = e \frac{\delta \hat H_0(\bk)}{\delta k^\alpha} = e v_F \sigma^{\alpha},
\ee
while the Matsubara Green's function has the form

\be
G(i\ve_n , \bk) = \frac12 \left[\frac{P_+(\bk)}{i \ve_n - v_F k  +\mu_1} + \frac{P_-(\bk)}{i \ve_n + v_F k  + \mu_1} \right],
\ee
and we introduced the projectors onto the conduction and the valence bands $P_{\pm} (\bk) = I \pm \hat \bk \cdot \boldsymbol{\sigma}$ with $\hat \bk \equiv \bk /k$. We emphasize again that we have only focused on the first node thus far; the contribution from the second node is obtained analogously.

Interestingly, the expression for $\chi(i\omega_1, i\omega_2)$ in the case of Weyl semimetals can be obtained exactly at $T=0$. Delegating the details of the calculation to the Supplemental Materials (SM)~\footnote{See Supplemental Materials for the details on the diagrammatic calculation of the CPGE coefficient and the interaction corrections to it.}, we present the answer:

\begin{widetext}
\be
\chi^{\alpha \beta \gamma}(i \omega_1, i \omega_2) = \frac{e^3}{48 \pi^2} \ve^{\alpha \beta \gamma}  \frac{\Omega^3 (\omega_2 - \omega_1) \ln\left( 4\mu_1^2 + \Omega^2  \right) + \omega_1^3 (\omega_2 + \Omega)\ln\left( 4\mu_1^2 + \omega_1^2   \right) - \omega_2^3 (\omega_1 + \Omega)\ln\left( 4\mu_1^2 + \omega_2^2   \right)}{\omega_1 \cdot \omega_2 \cdot \Omega}, \label{Eq:chianswer}
\ee
\end{widetext}
where $\ve^{\alpha \beta \gamma}$ is the fully antisymmetric Levi-Civita tensor. Equation~(\ref{Eq:chianswer}) along with Eq.~(\ref{Eq:currentgeneral}) is the first important result of our work, which describes the second-order response to  external  electric fields at arbitrary frequencies.

To obtain the injection current, we need to perform the analytic continuation to real frequencies, $i \omega_{1,2} \to \omega_{1,2} + i0,$ and set $\omega_1 = \omega + \Omega,$ $\omega_2 = - \omega$ with $\Omega \to 0$~\footnote{The analytic continuation requires some extra care. Matsubara frequencies $\omega_1$ and $\omega_2$ must be taken either both positive or both negative. Otherwise, the real-frequency answer will be incorrect.}:

\be
j^\gamma(\Omega) = - \frac1{12\pi} \frac{e^3}{\Omega} \ve^{\alpha \beta \gamma}  E^{\alpha}_\omega E^{\beta}_{-\omega}\Theta(\omega - 2|\mu_1|).
\ee
In the time domain, the $\Omega \to 0$ limit exactly corresponds to Eq.~(\ref{Eq:CPGEquant}) with $C=1$ and the CPGE coefficient given by

\be
\beta_0 = i \, \frac{\pi e^3}{3 h^2} \Theta(\omega - 2|\mu_1|). \label{Eq:beta_0}
\ee
Here we explicitly restored the Planck's constant $h=2\pi \hbar$ for clarity. 

The contribution from the second Weyl point has a similar form, but with the opposite sign due to different chirality, and $\mu_2$ instead of $\mu_1$ in the Heaviside step function. Consequently, in the frequency range $2 |\mu_1| < \omega < 2|\mu_2|,$ the CPGE in a noninteracting Weyl system becomes truly quantized and does not depend on the material-specific parameters such as the Fermi velocity, the exact position of the chemical potential, or the distance between the nodes, and is given by Eq.~(\ref{Eq:CPGEquant}). As we show below, the perfect quantization breaks down in the presence of interactions.

{\it Interaction corrections to the CPGE: the Hubbard potential. ---} Now we demonstrate by an explicit calculation that the electron-electron interactions destroy the quantization of the CPGE. As an example, we start with the Hubbard interaction and consider the static Coulomb potential later.

Generally (pseudospin conserving) electron-electron interaction is described by a Hamiltonian of the form

\begin{align}
H_{\text{int}} = \frac12 \sum_{i,j=1}^{2}\sum_{\bk, \bp, \bq} \psi^\dagger_{\bk - \bq,i,s} \psi_{\bk,i,s} \psi^\dagger_{\bp + \bq,j,s'} \psi_{\bp,j,s'} V(\bq) + \nonumber \\ + \frac12 \sum_{i=1}^{2}\sum_{\bk, \bp, \bq} \psi^\dagger_{\bk - \bq,i,s} \psi_{\bk,\bar i,s} \psi^\dagger_{\bp + \bq,\bar i,s'} \psi_{\bp,i,s'} V(\bK_0). \label{Eq:Hintgeneral}
\end{align}
We explicitly write down the summation over the  nodal indices $i, j$, and the summation over the pseudospin indices $s, s'$ is implied. Symbol $\bar i$ stands for the node different from node $i$ and $\bK_0$ is the distance between the nodes in momentum space (we assume $K_0 \gg p,k,q$).

The first term in Eq.~(\ref{Eq:Hintgeneral}) stands for the intranodal scattering, while the second one describes the scattering between the nodes, and we neglect processes that do not conserve the number of particles within each  node separately  (since they also violate momentum conversation).

\begin{figure}
  \centering
  \caption{ First-order self-energy [(a)-(d)] and vertex [(e)-(f)] corrections. Solid and dashed lines correspond to the Green's function of the first and second node, respectively. Diagrams (a), (c), and (e) describe the intranodal processes, while (b), (d), and (f) stand for the internodal scattering (important only for the Hubbard interaction).}
  \includegraphics[width=1.\columnwidth]{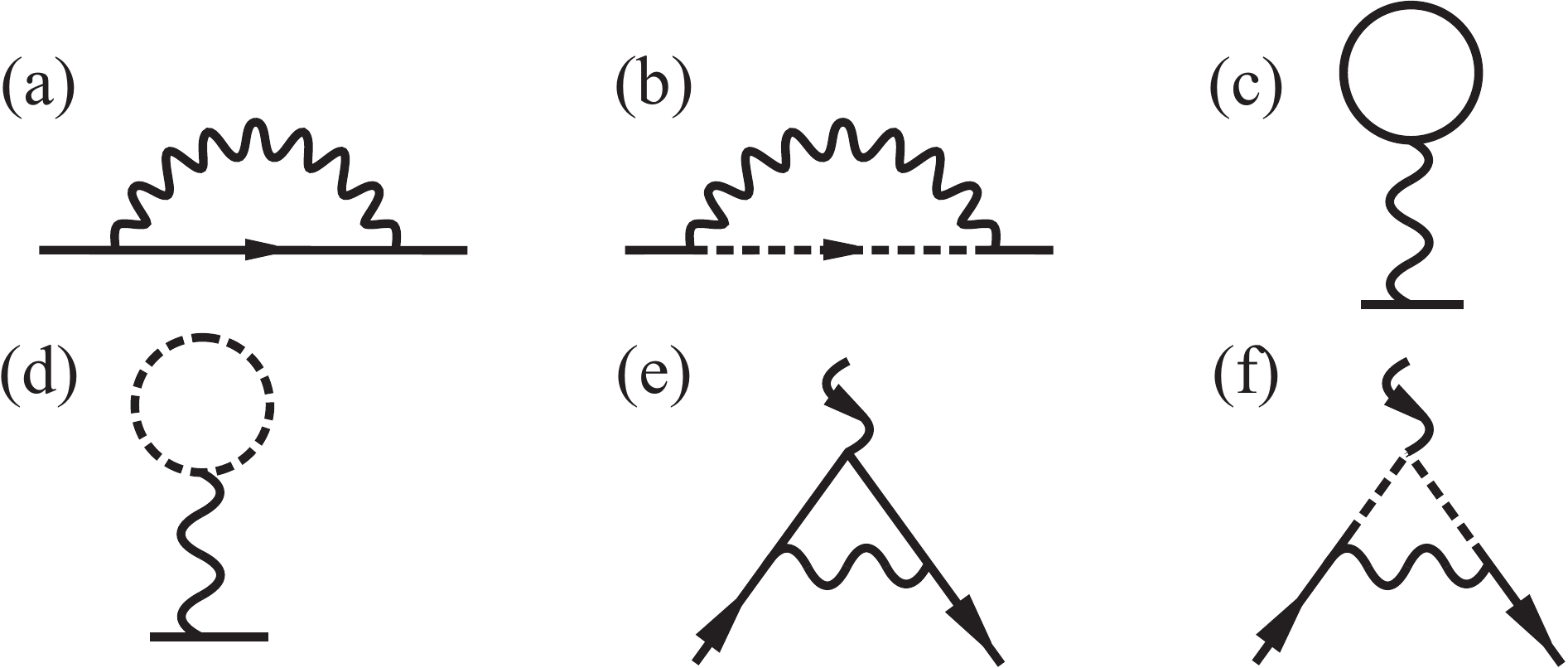}
  \label{Fig:selfenergyvertex}
\end{figure}

To the first order in interaction, corrections to photocurrent are given by the self-energy and vertex corrections shown diagrammatically in Fig.~\ref{Fig:selfenergyvertex}. Solid and dashed lines correspond to the electron propagators of the first and the second nodes, respectively.

In the case of the Hubbard potential, $V(\bq) = -\lambda$, the self-energy diagrams are just proportional to the total number of holes in the first node, $N_h$, or the number of electrons in the second node, $N_e$: $\Sigma^{(a)} = -\Sigma^{(c)}/2 = - \lambda N_h/2$, $\Sigma^{(b)} = -\Sigma^{(d)}/2 = \lambda  N_e/2.$
Taken together, these corrections only renormalize chemical potential, $ \delta \mu = - \sum_i \Sigma^{(i)} = \lambda \left( N_e - N_h  \right)/2$, which, in turn, shifts the range of frequencies where the CPGE is observed. This correction does not change the quantized value of the CPGE itself.

The vertex corrections, on the contrary, have a more profound effect and destroy the quantization of the CPGE. The correction to the CPGE coefficient $\beta_0$, Eq.~(\ref{Eq:beta_0}), due to the intranodal interaction is given by diagram~\ref{Fig:selfenergyvertex}(e) and equals [after the summation over all three current vertices in Eq.~(\ref{Eq:chi})]

\begin{multline}
\delta \beta^{(1)}(\omega) = - \beta_0 \cdot \frac{\lambda  }{24 \pi^2 v_F^3}  \times \\ \times \left( 6 v_F^2 \Lambda^2 - 6 \mu_1^2 - \omega^2 \ln \frac{|\omega^2 - 4\mu_1^2|}{4 v_F^2 \Lambda^2 - \omega^2}  \right),
\end{multline}
where $\Lambda$ is the high-momentum ultraviolet (UV) cutoff.

The strong UV divergence of this result is cured once we take the internodal scattering into account. Indeed, the short-ranged nature of the Hubbard interaction allows for the corrections shown in Fig.~\ref{Fig:selfenergyvertex}(f), $\delta\beta^{(2)}(\omega)$, which contributes with the overall opposite sign due to the opposite chirality of the second node. Hence, after adding up both intra- and internodal contributions, we obtain the total correction to the CPGE coefficient (see SM for details)

 \begin{multline}
\delta \beta(\omega)  = \delta \beta^{(1)}(\omega) + \delta \beta^{(2)}(\omega) = \\ = - \beta_0 \cdot \frac{\lambda  }{24 \pi^2 v_F^3} \left( 6 \mu_2^2 - 6 \mu_1^2 - \omega^2 \ln \left| \frac{\omega^2 - 4\mu_1^2}{\omega^2 - 4\mu_2^2} \right| \right). \label{Eq:deltabetaHubbard}
\end{multline}

We see that the first-order interaction correction is free of the UV divergencies, but is non-zero and has a characteristic frequency dependence.

{\it Interaction corrections to the CPGE: the Coulomb potential. ---} The whole analysis for the Coulomb potential is similar to that for the Hubbard interaction, with few important differences which we highlight below. The static screened Coulomb interaction that we focus on is given by Eq.~(\ref{Eq:Hintgeneral}) with 

\be 
V(\bq) = \frac{4\pi e^2}{\ve_0 (q^2 + q_0^2)}
\ee 
where $e$ is the electron's charge, $\ve_0$ is the dielectric constant due to core electrons, and $q_0$ is the Thomas-Fermi wave vector, respectively. The latter can be expressed through the fine-structure constant and the density of states at the Fermi level~\cite{LvZhang2013}. We, however, keep it an independent parameter for the purpose of generality, so that the interaction has the same form as the Yukawa potential.

\begin{figure}
  \centering
  \caption{ The dependence of function $F$, Eq.~(\ref{Eq:deletabetaYukawa}), on $q_0$ at $|\mu|/\omega = 0.00, \, 0.40$, and $0.43$ for the cases of the soft-cutoff and the dimensional regularizations. At $v_F q_0 \gg \omega,$ all curves approach $F=0$. }
  \includegraphics[width=1.\columnwidth]{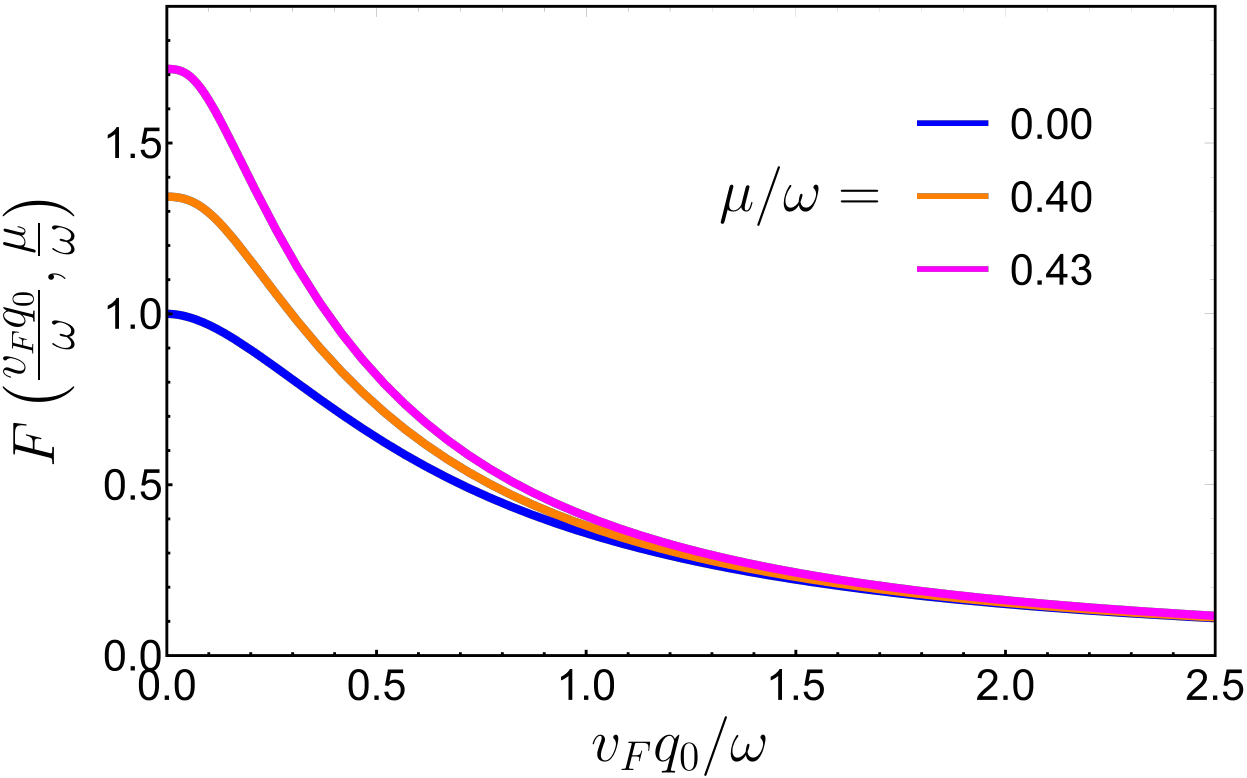}
  \label{Fig:Ffunction}
\end{figure}

Because of the long-ranged nature of the Coulomb interaction, one can focus on the correction due to the intranodal processes described by the first term in Eq.~(\ref{Eq:Hintgeneral}) only, while the contribution from the internodal scattering can be shown to be parametrically small. The correction is given by diagrams~\ref{Fig:selfenergyvertex}(a) and~\ref{Fig:selfenergyvertex}(e), and diagram~\ref{Fig:selfenergyvertex}(c) describes the $q=0$ component of the Coulomb interaction which is cancelled by the positive background. It can be straightforwardly shown that both the self-energy and vertex corrections to the CPGE coefficient $\beta$ are logarithmically UV divergent, see SM. The total answer, however, does not explicitly depend on the UV cutoff $\Lambda$ and is given by

\be 
\delta \beta = \beta_0 \frac{e^2}{\pi v_F \ve_0} F\left(\frac{v_F q_0}{\omega} ,  \frac{|\mu_1|}{\omega}  \right). \label{Eq:deletabetaYukawa}
\ee 
The function $F(x,y)$ is a smooth function independent of $\Lambda$, which is shown in Fig.~\ref{Fig:Ffunction} and with the exact expression given in SM. It turns out, however, that the particular form of $F(x,y)$ is sensitive to the regularization procedure. Thus, the answer obtained within the hard-cutoff (hc) regularization (which effectively cuts off the electron spectrum beyond the UV momentum scale $\Lambda$) is different from that obtained by the soft-cutoff (sc) and dimensional regularization (dr), and they all are related according to

\be  
F^{\text{sc}}(x,y)  = F^{\text{dr}}(x,y) = F^{\text{hc}}(x,y) -1. \label{Eq:FscFdrFhc}
\ee 
This peculiar result is similar to what happens with the interaction correction to the optical conductivity in graphene. The origin of the discrepancy is rooted in the way that different regularization procedures account for the high-energy (of order $v_F \Lambda$) states. The hard-cutoff scheme completely neglects the high-energy contribution and, as a result, violates the Ward-Takahashi identity. Hence the answer obtained within this procedure is only qualitatively correct.

The interaction corrections~(\ref{Eq:deltabetaHubbard}) and (\ref{Eq:deletabetaYukawa}), along with the general response function for the noninteracting system~(\ref{Eq:chianswer}), are the main calculational results of this work.   In graphene, the interaction corrections seem experimentally to be small~\cite{Nairetal2008}, but in the present case we expect the interaction corrections to be significant, unless the effective dielectric constant is rather large, and potentially observable.

{\it Conclusions. ---} In conclusion, using the Hubbard and the static Coulomb interactions as examples, we have shown that the interactions destroy the quantization of the CPGE. We have found that, in the case of the Coulomb interaction, the correction depends on the way one regularizes the contribution from the high-energy states, leading to the wrong result if the hard cutoff is used. This result is similar to that for the interaction correction to optical conductivity in graphene. Unlike graphene, however, where the quantization of optical conductivity even in a  noninteracting system is not protected by topology, the quantization of the CPGE in noninteracting Weyl semimetals is tight to the monopole strength of the Weyl nodes. Hence, our result implies that, since the topological charge of the node remains unchanged, the interactions change the relation between the CPGE injection current and the nodal strength. It may be possible to observe the interaction effects on the frequency dependence of the plateau in the photocurrent, especially if the effects of disorder can be minimized by a short pulse or a difference-frequency-generation approach~\cite{fern2019difference}.  We expect the same qualitative results to hold for the higher-order nodal materials~\cite{Flickeretal2018}, though we leave an explicit calculation in this case for a future study.

{ \it Acknowledgments. ---} We thank Fernando de Juan, Adolfo Grushin, and Daniel Parker for discussions and collaboration on related projects. We also appreciate comments from J\"org Schmalian, Inti Sodemann, and Igor Herbut. This work was supported by the Quantum Materials program at LBNL, funded by the US Department of Energy under Contract No. DE-AC02-05CH11231 (V.~K. and J. E. M.), the National Science Foundation under Grant No. DMR-1918065 (A. A.), and a Simons Investigatorship (J. E. M.).

\bibliography{cpge}

\newpage

\begin{widetext}

\begin{center}
\textbf{\large Supplemental Materials for ``Interactions remove the quantization of the chiral photocurrent at Weyl points''} 
\end{center}
\setcounter{equation}{0}
\setcounter{figure}{0}
\setcounter{table}{0}

\makeatletter
\renewcommand{\theequation}{S\arabic{equation}}
\renewcommand{\thefigure}{S\arabic{figure}}
\renewcommand{\thetable}{S\Roman{table}}

This Supplemental Material consists of three sections. In Section~I, we use the method of Feynman diagrams to rederive the quantized result for the circularly polarized photogalvanic effect (CPGE) in noninteracting Weyl semimetals obtained in Ref.~\cite{deJuanGrushinMorimotoMoore2017}. Then, we calculate the first-order interaction corrections due to the Hubbard interaction in Section~II and the screened Coulomb potential in Section~III.

\section{I. Quantized result in the absence of interactions \label{App:non-interacting}}

\begin{figure}[b]
  \centering
  \caption{ Diagrams contributing to the quantized circular photogalvanic effect in the absence of interactions.}
  \includegraphics[width=.7\columnwidth]{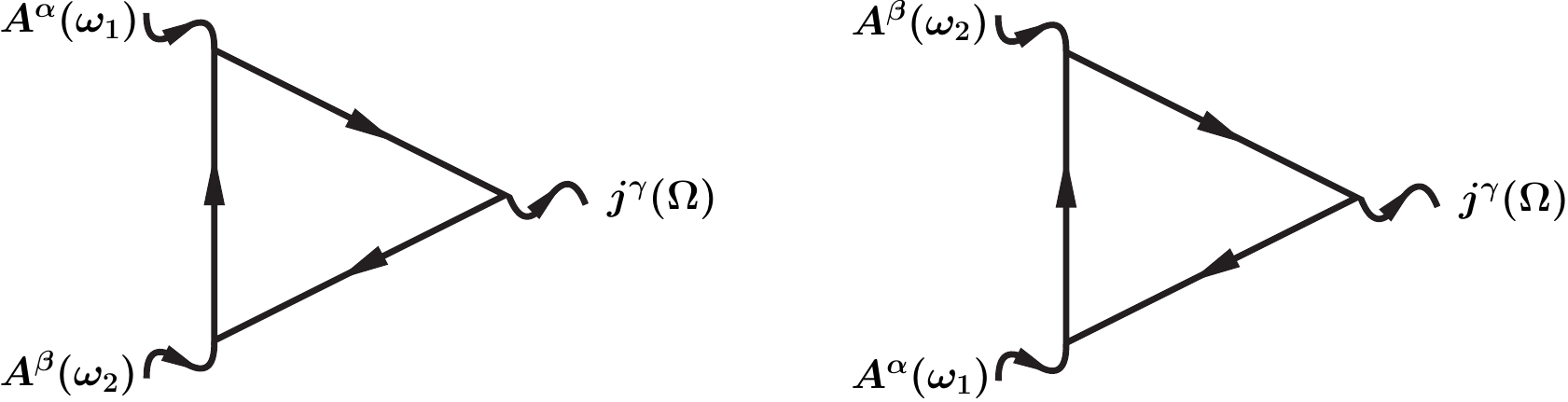}
  \label{SMFig:bareresponse}
\end{figure}

In the absence of interactions, the second-order current response to the external electric field is given by the two diagrams shown in Fig.~\ref{SMFig:bareresponse}, see Refs.~\cite{JoaoLopes2018} and~\cite{ParkeretalDiagrams2019}. These diagrams describe the three-current correlation function, where the current operator $\hat j^{\alpha}$ for a Weyl fermion, Eq.~(\ref{Eq:H0}), is given by
\be
\hat j^{\alpha} = e \frac{\delta \hat H(\bk)}{\delta k^\alpha} = \pm e v_F \sigma^{\alpha}.
\ee
Here, $\sigma^{\alpha}$ are Pauli matrices, and ``$\pm$'' sign reflects different chiralities of the nodes. More specifically, the expression for the photocurrent reads as

\be
j^\gamma(\Omega) = \frac{1}{\omega_1 \omega_2} \left[ \chi_1^{\alpha \beta \gamma}(\omega_1, \omega_2) +  \chi_2^{\alpha \beta \gamma}(\omega_1, \omega_2) \right] E^{\alpha} (\omega_1) E^{\beta}(\omega_2), \label{SMEq:currentgeneral}
\ee
with $\Omega \equiv \omega_1 + \omega_2,$ and the contributions $\chi_1$ and $\chi_2$ are given by the left and right diagrams in Fig.~\ref{SMFig:bareresponse}, correspondingly. The analytical expressions for $\chi_{1,2}$ can be straightforwardly calculated in the Matsubara formalism. For example, the expression for the first diagram reads as

\be
\chi^{\alpha \beta \gamma}_1(i \omega_1, i\omega_2) = T\sum_{\ve_n} \int \frac{d^3 k}{(2\pi)^3} \text{tr} \, \left[ j^{\alpha} G(i\ve_n - i\omega_1, \bk) j^{\beta} G(i\ve_n - i \Omega, \bk) j^{\gamma} G(i\ve_n, \bk)\right].
\ee
where $\ve_n = \pi T (2n+1)$ and $T$ is the temperature. The expression for the second diagram can be simply obtained as

\be
\chi^{\alpha \beta \gamma}_2(i \omega_1, i\omega_2) = \chi^{ \beta \alpha \gamma}_1(i \omega_2, i\omega_1).
\ee

If interactions are absent, one can calculate the contributions from each node separately. The Green's function for the first Weyl node is given by

\be
G(i\ve_n , \bk) = \frac12 \left[\frac{P_+(\bk)}{i \ve_n - v_F k -|\mu_1|} + \frac{P_-(\bk)}{i \ve_n + v_F k -|\mu_1|} \right],
\ee
where we introduced the projectors onto the conduction (``+'') and the valence (``-'') bands

\be
P_{\pm} (\bk) = I \pm \hat \bk \cdot \boldsymbol{\sigma}, \qquad \hat \bk \equiv \bk/k,
\ee
and we chose chemical potential to be negative for definiteness, $\mu_1 < 0$.

At zero temperature, one has $T \sum_{\ve_n}\ldots \to \int (d\ve/2\pi) \ldots$

The integral over the intermediate frequency $\ve$ can be evaluated exactly yielding

\begin{align}
\chi_1^{\alpha \beta \gamma}(i \omega_1, i \omega_2) &= \frac{e^3 v_F^3}8 \int \frac{d^3 k}{(2\pi)^3} \Theta(v_F k -|\mu_1|) \tr \left[ \frac{\sigma^{\alpha} P_-(\bk) \sigma^{\beta} P_+(\bk) \sigma^{\gamma} P_+(\bk)}{(2v_F k - i \omega_1)(2 v_F k + i \omega_2)} + \frac{\sigma^{\alpha} P_+(\bk) \sigma^{\beta} P_+(\bk) \sigma^{\gamma} P_-(\bk)}{(2v_F k + i \Omega)(2 v_F k + i \omega_1)}   - \right. \nonumber  \\  &   - \frac{\sigma^{\alpha} P_+(\bk) \sigma^{\beta} P_-(\bk) \sigma^{\gamma} P_-(\bk)}{(2v_F k - i \omega_2)(2 v_F k + i \omega_1)} + \frac{\sigma^{\alpha} P_+(\bk) \sigma^{\beta} P_-(\bk) \sigma^{\gamma} P_+(\bk)}{(2v_F k - i \omega_2)(2 v_F k - i \Omega)}  - \frac{\sigma^{\alpha} P_-(\bk) \sigma^{\beta} P_+(\bk) \sigma^{\gamma} P_-(\bk)}{(2v_F k + i \omega_2)(2 v_F k + i \Omega)}  -  \nonumber \\ &\left. - \frac{\sigma^{\alpha} P_-(\bk) \sigma^{\beta} P_-(\bk) \sigma^{\gamma} P_+(\bk)}{(2v_F k - i \omega_1)(2 v_F k - i \Omega)} \right], \label{SMEq:chi1overepsilon}
\end{align}
with $\Theta(x)$ the Heaviside step function.

Averaging over the directions of $\bk$ gives

\be
\langle  \tr \left[  \sigma^{\alpha} P_+(\bk) \sigma^{\beta} P_+(\bk) \sigma^{\gamma}P_-(\bk)  \right] \rangle_{\hat \bk} = \langle  \tr \left[  \sigma^{\alpha} P_+(\bk) \sigma^{\beta} P_-(\bk) \sigma^{\gamma}P_-(\bk) \right] \rangle_{\hat \bk} = \frac83 i \ve^{\alpha \beta \gamma}, \label{SMEq:corr0}
\ee
where $\ve^{\alpha \beta \gamma}$ is the fully antisymmetric Levi-Civita tensor. All other correlators can be simply obtained by the permutation of indices.

Finally, collecting all the terms in Eq.~(\ref{SMEq:chi1overepsilon}) together and performing the integration over $\bk$, one finds the general analytical expression 

\be
\chi_1^{\alpha \beta \gamma}(i \omega_1, i \omega_2) = \frac{e^3}{48 \pi^2} \cdot \ve^{\alpha \beta \gamma} \cdot \frac{\Omega^3 (\omega_2 - \omega_1) \ln\left( 4\mu_1^2 + \Omega^2  \right) + \omega_1^3 (\omega_2 + \Omega)\ln\left( 4\mu_1^2 + \omega_1^2   \right) - \omega_2^3 (\omega_1 + \Omega)\ln\left( 4\mu_1^2 + \omega_2^2   \right)}{\omega_1 \cdot \omega_2 \cdot \Omega}.
\ee

To find the actual physical response, we need to perform analytic continuation of the above expression to real frequencies. This is a subtle procedure and must be carried out with extra care. In particular, in order to obtain a physically meaningful result, we must take  $\omega_1$ and $\omega_2$ either both positive or both negative. This statement can be directly checked by using the Keldysh technique~\cite{JoaoLopes2018}. Choosing $\omega_{1,2} > 0$ for definiteness, analytic continuation is  performed by taking

\be
i \omega_{1,2} \to \omega_{1,2} + i\delta, \qquad \delta \to +0. \label{SMEq:analyticalcontinuation}
\ee

The logarithms then transform according to

\be
\ln\left[4\mu_1^2 + \omega^2 \right] \to \ln\left[4\mu_1^2 - (\omega + i \delta)^2 \right] = \ln |4\mu_1^2 - \omega^2|- i \, \pi\,  \text{sign}( \omega) \,  \Theta\left(|\omega| - 2|\mu_1|\right).
\ee

The injection current which we are interested in here corresponds to the specific choice $\omega_1 = \omega + \Omega$, $\omega_2 = - \omega$, with $\Omega \to 0$. After the analytic continuation, we find in this limit

\be
\chi_1^{\alpha \beta \gamma}( \omega + \Omega, -\omega) =  - \frac{e^3}{24 \pi} \cdot \ve^{\alpha \beta \gamma} \cdot \frac{\omega^2}{\Omega} \cdot \Theta (\omega - 2|\mu_1|). \label{SMEq:chi1answer}
\ee
The same contribution comes from the second diagram in Fig.~\ref{SMFig:bareresponse}. Collecting them together, we find for the current from Eq.~(\ref{Eq:currentgeneral})

\be
j^\gamma(\Omega) = - \frac1{12\pi} \frac{e^3}{\Omega} \ve^{\alpha \beta \gamma}  E^{\alpha} (\omega + \Omega) E^{\beta}(-\omega) \Theta(\omega - 2|\mu_1|).
\ee
In the time domain, this corresponds to

\be
\frac{d}{dt} j^{i} = \beta_0(\omega) \left[ \bE_\omega \times \bE_{-\omega} \right]^{i}, \label{SMEq:dj/dt}
\ee
with

\be
\beta_0(\omega) = \frac{i e^3}{12 \pi} \Theta(\omega - 2|\mu_1|), \label{SMEq:beta0}
\ee
in full agreement with Ref.~\onlinecite{deJuanGrushinMorimotoMoore2017}.

Result~(\ref{SMEq:beta0}) was obtained for the first Weyl node with the chemical potential $\mu_1$. Analogously, one obtains for the second node  

\be
\beta_{0}^{(2)}(\omega) = - \frac{i e^3}{12 \pi} \Theta(\omega - 2|\mu_2|) . 
\ee
Consequently, in the frequency range $2|\mu_1| < \omega < 2 |\mu_2|$, only the first node contributes to the CPGE, while the contribution from the second node vanishes due to Pauli blocking. This conclusion (at least, to the leading order) holds even in the presence of static interactions.

Result~(\ref{SMEq:dj/dt})-(\ref{SMEq:beta0}) can be readily derived directly from Eq.~(\ref{SMEq:chi1overepsilon}). In fact, it is straightforward to check that only the first term in Eq.~(\ref{SMEq:chi1overepsilon}) contributes to the injection current. Indeed, performing analytic continuation~(\ref{SMEq:analyticalcontinuation}), setting $\omega_1 = \omega + \Omega,$ $\omega_2 = -\omega$, and focusing on the imaginary part of the expression, we find

\be 
\frac1{(2v_F k - i\omega_1) (2v_F k + i\omega_2)} \to\pi i \left[    \frac{\delta(2 v_F k - \omega - \Omega)}{2v_F k - \omega } - \frac{\delta (2v_F k - \omega)}{2 v_F k - \omega - \Omega} \right] \approx \frac{2 \pi i \delta (2v_F k - \omega)}{\Omega}, \label{SMEq:deltafunction}
\ee
where we also used $\Omega \to 0$ in the last equality. Performing now trivial integration over $k$, we immediately obtain Eq.~(\ref{SMEq:chi1answer}). This more straightforward approach will allow us to significantly simplify the calculation of the interaction correction in Sec.~III.

\begin{figure}[b]
  \centering
  \caption{ Diagrammatic representation of the scattering processes described by Eq.~(\ref{SMEq:Hintgeneral}). Solid lines correspond to the Green's function of the first node, dashed lines are for the second node, wavy lines stand for interaction. Diagrams (a)-(c) describe the intranodal scattering, (d)-(e) are for internodal processes. The processes that do not conserve the number of particles within each node separately are not allowed by the momentum conservation.   }
  \includegraphics[width=1.\columnwidth]{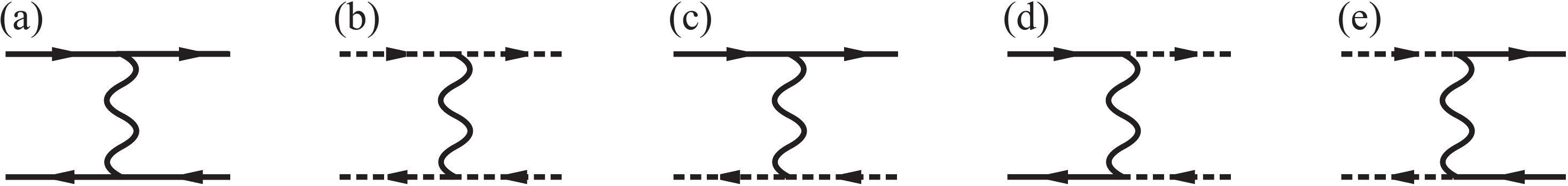}
  \label{SMFig:scattering}
\end{figure}

\section{II. Interaction corrections to the quantized CPGE: the Hubbard interaction \label{App:Hubbard}}

Now we consider the perturbative corrections originating from finite electron-electron interaction. The interaction Hamiltonian is given by Eq.~(\ref{Eq:Hintgeneral}):

\be
H_{\text{int}} = \frac12 \sum_{i,j=1}^{2}\sum_{\bk, \bp, \bq} \psi^\dagger_{\bk - \bq,i,s} \psi_{\bk,i,s} \psi^\dagger_{\bp + \bq,j,s'} \psi_{\bp,j,s'} V(\bq) + \frac12 \sum_{i=1}^{2}\sum_{\bk, \bp, \bq} \psi^\dagger_{\bk - \bq,i,s} \psi_{\bk,\bar i,s} \psi^\dagger_{\bp + \bq,\bar i,s'} \psi_{\bp,i,s'} V\left(\bq + (-1)^{i+1} \bK_0 \right). \label{SMEq:Hintgeneral}
\ee
The first term in the above expression describes the intranodal scattering processes, while the second one stands for the internodal scattering. The summation over the nodal indices $i,j=1,2$ is explicit here, while the summation over the pseudospin indices $s, s'$ is implied. $\bK_0$ in the above expression is the separation between the nodes in momentum space, and $\bar i$ designates the node different from $i$. We assume that the nodes are well separated in momentum space, consequently, the processes that do not conserve the number of particles within each node separately violate the momentum conservation and hence are not allowed. Furthermore, this assumptions implies that $q\ll K_0$, and one can substitute $V(\bq \pm \bK_0) \to V(\bK_0)$ in the second term. The scattering processes described by Hamiltonian~(\ref{SMEq:Hintgeneral}) are shown diagrammatically in Fig.~\ref{SMFig:scattering}.

In this section, we consider the case of the constant Hubbard interaction with strength $\lambda$, $V(\bq) = -\lambda$ (positive $\lambda$ corresponds to the attractive interaction). We only consider the first-order self-energy and vertex corrections.

\subsection{Self-energy corrections}

\begin{figure}[b]
  \centering
  \caption{ First-order contributions to self-energy. Diagrams (b) and (d), describing the internodal scattering, are only important in the case of the Hubbard interaction, and can be neglected in the case of the Coulomb potential.}
  \includegraphics[width=.8\columnwidth]{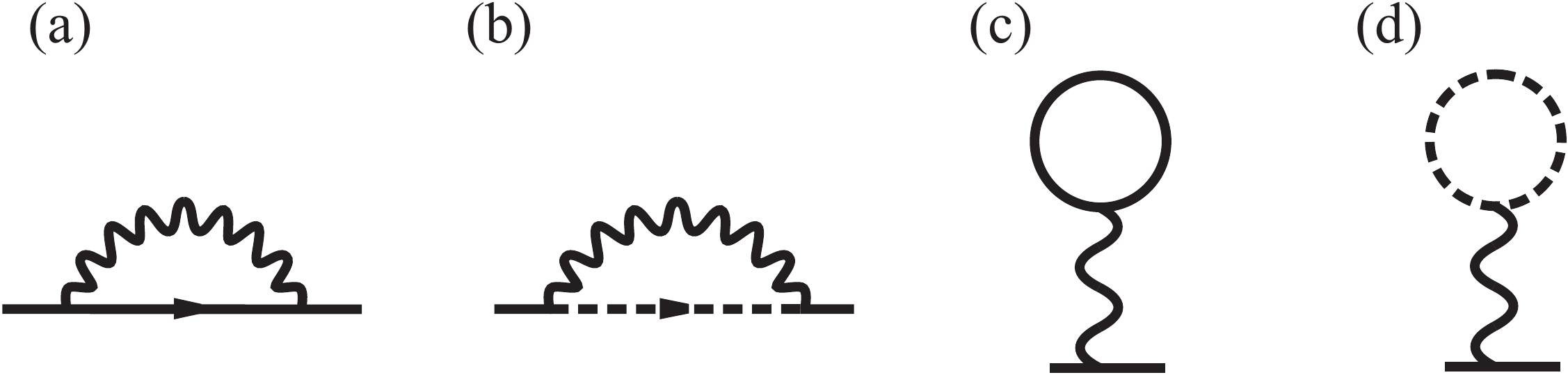}
  \label{SMFig:self-energy}
\end{figure}

The contributions to the first-order self-energy correction are shown diagrammatically in Fig.~\ref{SMFig:self-energy}. We consider the Hubbard interaction to be truly short-ranged in a sense that it allows for scattering between different Weyl nodes (solid lines correspond to the first Weyl point, and the dashed lines represent the second Weyl point). Hence, the second term in Eq.~(\ref{SMEq:Hintgeneral}) cannot be neglected and should also be taken into account. The expression for diagram~\ref{SMFig:self-energy}~(a) reads as

\be
\Sigma^{(a)} = \lambda T\sum_{\ve'_n} \sum_{\bk} G(i \ve_n, \bk) = -\frac{\lambda}2 \sum_{\bk} \left[ 1 - \Theta(v_F k - |\mu_1|) \right] = - \frac{\lambda}2 N_h,
\ee
where $N_h > 0$ is the number of holes below the Weyl point in the first node. When calculating the integral over the intermediate energies $\ve'$ in the above expression, we took the half-sum of contours closed in the upper and the lower half-planes.

Analogously, the contribution from diagram~\ref{SMFig:self-energy} (b) equals

\be
\Sigma^{(b)} = \frac{\lambda}2 N_e,
\ee
with $N_e > 0$ the number of electrons above the Weyl point in the second node.

Finally, the contributions from diagrams~\ref{SMFig:self-energy} (c) and (d) equal

\be
\Sigma^{(c)} = -2 \Sigma^{(a)}, \qquad \Sigma^{(d)} = -2 \Sigma^{(b)},
\ee
resulting in the total self-energy

\be
\Sigma = \Sigma^{(a)} + \Sigma^{(b)} + \Sigma^{(c)} + \Sigma^{(d)} = -\frac{\lambda}2 \left( N_e - N_h  \right).
\ee
This self-energy simply shifts the chemical potential according to

\be
\delta \mu = - \Sigma = \frac{\lambda}2 \left( N_e - N_h  \right). \label{SMEq:deltamu}
\ee
It does not change the CPGE coefficient $\beta$, and only modifies the frequency range  where the quantization is observed.

\begin{figure}[b]
  \centering
  \caption{ First-order vertex corrections contributing to the renormalization of the CPGE coefficient. Diagram (a) describes intranodal processes, while (b) stands for the internodal scattering. In the case of the Hubbard interaction, both diagrams must be taken into account. In the case of the Coulomb interaction, diagram (b) is parametrically smaller than (a) and can be neglected.}
  \includegraphics[width=.55\columnwidth]{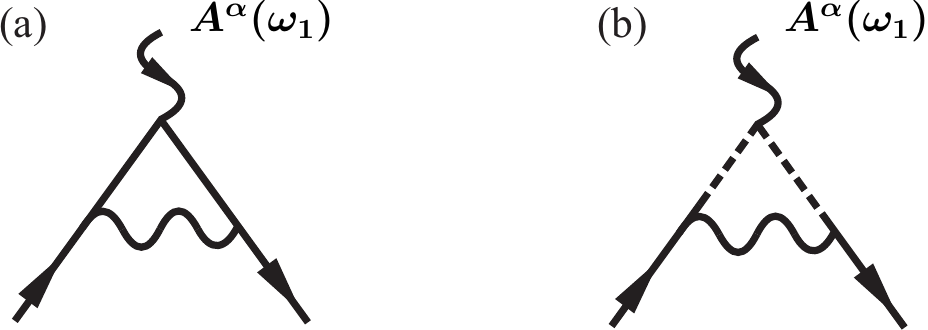}
  \label{SMFig:vertex}
\end{figure}

\subsection{Vertex corrections}

The first-order vertex corrections are shown in Fig.~\ref{SMFig:vertex}. Focusing on the vertex $\alpha$ with the external Matsubara frequency $\omega_1$ for definiteness, the first diagram reads as 

\be
v_F\sigma^\alpha \to \lambda v_F T\sum_{\ve'} \sum_{\bp} G(i \ve', \bp) \sigma^{\alpha} G(i\ve' - i\omega_1, \bp) = - \frac{\lambda \sigma^{\alpha}}{48 \pi^2 v_F^2} \left[4 v_F^2 \Lambda^2 - 4\mu_1^2 + \omega_1^2 \ln\frac{\omega_1^2 + 4\mu_1^2}{\omega_1^2 + 4v_F^2\Lambda^2}  \right],
\ee
where $\Lambda$ is the UV momentum cutoff.

The contribution from the second diagram is analogous but has an overall opposite sign due to the opposite chirality of the second node and with $\mu_1^2 \to \mu_2^2$:

\be
v_F\sigma^\alpha \to   \frac{\lambda \sigma^{\alpha}}{48 \pi^2  v_F^2} \left[4  v_F^2 \Lambda^2 - 4\mu_2^2 + \omega_1^2 \ln\frac{\omega_1^2 + 4\mu_2^2}{\omega_1^2 + 4 v_F^2\Lambda^2}  \right].
\ee
Adding these two contributions together, we find that the vertex with $\sigma^{\alpha}$ (and $\omega_1$) is renormalized according to

\be
\sigma^{\alpha} \to - \frac{\lambda \sigma^{\alpha}}{48 \pi^2 v_F^3}\left[4\mu_2^2 - 4\mu_1^2 + \omega_1^2 \ln\frac{\omega_1^2+4\mu_1^2}{\omega_1^2 + 4\mu_2^2}  \right],
\ee
which is finite and does not contain the UV cutoff anymore.

Analogous contributions are obtained for vertices with $\sigma^{\beta}$ (frequency $\omega_2$) and $\sigma^{\gamma}$ (frequency $\Omega = \omega_1 + \omega_2$). Finally, we note that the contributions to the interaction correction from both diagrams in Fig.~\ref{SMFig:bareresponse} are the same. Summing up the contributions from all three vertices, performing the analytic continuation $i\omega_{1,2} \to \omega_{1,2} + i\delta$, and setting $\omega_1 = - \omega_2 = \omega$, we find that the CPGE coefficient acquires a correction $\delta \beta$:

 \be
\delta \beta(\omega)   = - \beta_0 \cdot \frac{\lambda  }{24 \pi^2 v_F^3} \left( 6 \mu_2^2 - 6 \mu_1^2 - \omega^2 \ln \left| \frac{\omega^2 - 4\mu_1^2}{\omega^2 - 4\mu_2^2} \right| \right).
\ee
in accordance with Eq.~(\ref{Eq:deltabetaHubbard}). In the last equation, we neglected the corrections to the chemical potentials, since they only change the frequency range where the bare (noninteracting) CPGE is non-zero.

\section{III. Interaction corrections to the quantized CPGE: the Coulomb interaction \label{App:Coulomb}}

The static screened Coulomb interaction is described by Hamiltonian~(\ref{SMEq:Hintgeneral}) with

\be
V(\bq) = \frac{4\pi e^2}{\ve_0(q^2 + q_0^2)}. \label{SMEq:Yukawapotential}
\ee
We assume that the nodes are well separated in momentum space, so the internodal scattering processes are suppressed and the second term in Eq.~(\ref{SMEq:Hintgeneral}), along with diagrams~\ref{SMFig:self-energy} (b), (d) and~\ref{SMFig:vertex} (b), can be neglected to the leading order. Indeed, realistically, the separation between the nodes $K_0$ and the UV cutoff $\Lambda$ up to which the spectrum of fermions can be considered as linear are related as $\Lambda \lesssim K_0$, so the leading-order contribution of the internodal scattering in the case of the Coulomb interaction is proportional to $\sim (\Lambda/K_0)^2 \ll 1$. Furthermore, diagram~\ref{SMFig:self-energy} (c) corresponds to the $q=0$ component of the Coulomb potential, which is cancelled by the positive background.

\subsection{Self-energy correction}

The lowest-order self-energy correction is given by diagram~\ref{SMFig:self-energy} (a):

\be
\Sigma(\ve, \bk) = - T\sum_{\ve'}\sum_{\bq} G(\ve',\bq) V(\bk - \bq) = \frac{ 2\pi e^2}{\ve_0} \left[ f_1(k) + (\bk \cdot \boldsymbol{\sigma})  f_2(k) \right], \label{SMEq:SigmaYukawa}
\ee
with

\begin{align}
f_1(k) & = \sum_{v_F p < |\mu_1|} \frac1{(\bk - \bp)^2+q_0^2}    = \frac1{8\pi^2 k}\int_0^{|\mu_1|/v_F} p \, dp \, \ln \frac{(p+k)^2+q_0^2}{(p-k)^2+q_0^2}, \nonumber \\ f_2(k) & = \frac1k \sum_{v_F p > |\mu_1|} \frac{\hat \bk \cdot \hat \bp}{(\bk - \bp)^2+q_0^2} = \frac1{4 \pi^2 k} \int_{|\mu_1|/v_F}^{\Lambda} p^2 \, dp \int_{-1}^{1}\frac{t \, dt}{p^2 + k^2 + q_0^2 - 2 p k t}, \label{SMEq:f1-f2}
\end{align}
where $t= \cos \theta$, and $\theta$ is the angle between $\bk$ and $\bp$. Again, performing the integration over $\ve'$, we took the half-sum of contours closed in the upper and lower half-planes.

The term with $f_1(k)$ can be viewed as the renormalization of the chemical potential plus the higher-order in $k$ corrections to the spectrum, while $f_2(k)$ describes the $k$-dependent renormalization of the Fermi velocity. To see the effect of these terms on the CPGE coefficient $\beta$, we refer to Eqs.~(\ref{SMEq:chi1overepsilon}) and~(\ref{SMEq:deltafunction}). It is straightforward to show that $f_1$ does not change the value of $\beta$, only the range of frequencies where the (noninteracting) CPGE is observed. The effect of the velocity renormalization, on the other hand, is significant and leads to the following leading-order correction 

\be 
\frac{\delta \beta_{\text{self-energy}}}{\beta_0} = - \frac{2 \pi e^2}{v_F \ve_0} \left[ 3 f_2 \left( \frac{\omega}{2v_F} \right) +  \frac{\omega}{2v_F}  f'_2 \left( \frac{\omega}{2v_F} \right)\right].  \label{SMEq:deltabSE}
\ee
While the first term in this expression merely comes from the $\sim 1/v_F^3$ dependence in Eq.~(\ref{SMEq:chi1answer}), the second term is more subtle and reflects the fact that the $k$-dependent correction to the Fermi velocity appears in the argument of the $\delta$-function in Eq.~(\ref{SMEq:deltafunction}). The argument $\omega/2v_F$ in $f_2$ and $f_2'$ is also due to the $\delta$-function in Eq.~(\ref{SMEq:deltafunction}) which fixes the value of $k$.

As expected, the correction to the Fermi velocity, $f_2(k)$, is logarithmically UV-divergent. However, as we will show below, this logarithmic divergence cancels once we take the vertex corrections into account.

\subsection{Vertex correction}

The vertex correction is given by diagram~\ref{SMFig:vertex} (a) and reads as (we take vertex $\alpha$ with frequency $\omega_1$ for definiteness)

\begin{align}
&\sigma^\alpha \to  - T\sum_{\ve'} \sum_{\bp} G(i \ve', \bp) \sigma^{\alpha} G(i\ve' - i\omega_1, \bp) V(\bp - \bk) = \nonumber \\ & = \frac{2\pi e^2}{\ve_0 v_F} \left\{ \sigma^{\alpha} f_3(i\omega_1, k) - (\hat \bk \cdot \boldsymbol{\sigma}) \sigma^{\alpha} (\hat \bk \cdot \boldsymbol{\sigma}) f_4(i\omega_1, k) + [\hat \bk \times \boldsymbol{\sigma}]^{\alpha} f_5(i\omega_1, k)   \right\}, \label{SMEq:deltasigma}
\end{align}
with

\begin{align}
f_3(i\omega, k) &=  \sum_{v_F p > |\mu_1|} \frac{v_F^2 p \left[ 3 - (\hat \bp \cdot \hat \bk)^2  \right]}{\left[(\bk - \bp)^2+q_0^2\right] \left[ 4 v_F^2 p^2 + \omega^2 \right]} = \frac{v_F^2}{4 \pi^2} \int_{|\mu_1|/v_F}^{\Lambda} \frac{p^3 \, dp}{4 v_F^2 p^2 + \omega^2} \int_{-1}^{1}\frac{(3-t^2) \, dt}{p^2 + k^2 + q_0^2 - 2 p k t}, \nonumber \\  f_4(i\omega, k) &= \sum_{v_F p > |\mu_1|} \frac{v_F^2 p \left[ 3(\hat \bp \cdot \hat \bk)^2 - 1 \right]}{\left[(\bk - \bp)^2+q_0^2\right] \left[ 4 v_F^2 p^2 + \omega^2 \right]} = \frac{v_F^2}{4 \pi^2} \int_{|\mu_1|/v_F}^{\Lambda} \frac{p^3 \, dp}{4 v_F^2 p^2 + \omega^2} \int_{-1}^{1}\frac{(3t^2 -1) \, dt}{p^2 + k^2 + q_0^2 - 2 p k t}, \nonumber \\  f_5(i\omega, k) &= \sum_{v_F p > |\mu_1|} \frac{ 2 v_F \omega (\hat \bp \cdot \hat \bk)}{\left[(\bk - \bp)^2+q_0^2\right] \left[ 4 v_F^2 p^2 + \omega^2 \right]} = \frac{\omega v_F}{2 \pi^2} \int_{|\mu_1|/v_F}^{\Lambda} \frac{p^2 \, dp}{4 v_F^2 p^2 + \omega^2} \int_{-1}^{1}\frac{t \, dt}{p^2 + k^2 + q_0^2 - 2 p k t}. \label{SMEq:f3-f5}
\end{align}

To render these expressions into the actual corrections to the CPGE, we need to further perform several simple steps, analogously to the case of the Hubbard interaction. First, we need to sum over all three vertices $\alpha$, $\beta,$ and $\gamma$ in Fig.~\ref{SMFig:bareresponse}. Second, we analytically continue from Matsubara frequencies according to Eq.~(\ref{SMEq:analyticalcontinuation}) setting $\omega_1 = \omega + \Omega$, $\omega_2 = -\omega$, $\Omega \to 0$. Finally, since only the first term from Eq.~(\ref{SMEq:chi1overepsilon}) contributes to the injection current, we observe that  Eq.~(\ref{SMEq:deltafunction}) pins momentum $k$ to $ k = \omega/2v_F$. Then, exploiting equalities  

\begin{align}
&\langle  \tr \left[ (\hat \bk \cdot {\boldsymbol \sigma})  \sigma^{\alpha} (\hat \bk \cdot {\boldsymbol \sigma}) P_-(\bk) \sigma^{\beta} P_+(\bk) \sigma^{\gamma}P_+(\bk)  \right] \rangle_{\hat \bk} = \langle  \tr \left[   \sigma^{\alpha} P_-(\bk) (\hat \bk \cdot {\boldsymbol \sigma}) \sigma^{\beta} (\hat \bk \cdot {\boldsymbol \sigma}) P_+(\bk) \sigma^{\gamma}P_+(\bk)  \right] \rangle_{\hat \bk}  = \nonumber \\ = - &\langle  \tr \left[  \sigma^{\alpha}  P_-(\bk) \sigma^{\beta} P_+(\bk) (\hat \bk \cdot {\boldsymbol \sigma}) \sigma^{\gamma} (\hat \bk \cdot {\boldsymbol \sigma}) P_+(\bk)  \right] \rangle_{\hat \bk} = i \langle  \tr \left[ [\hat \bk \times {\boldsymbol \sigma}]^{\alpha} P_-(\bk) \sigma^{\beta} P_+(\bk) \sigma^{\gamma}P_+(\bk)  \right] \rangle_{\hat \bk} = \nonumber \\ =-i&\langle  \tr \left[ \sigma^{\alpha} P_-(\bk) [\hat \bk \times {\boldsymbol \sigma}]^{\beta} P_+(\bk) \sigma^{\gamma}P_+(\bk)  \right] \rangle_{\hat \bk} = -\frac{8 i}{3} \ve^{\alpha \beta \gamma}, \label{SMEq:corr1}
\end{align}
we find for the overall vertex correction

\be
\frac{\delta \beta_{\text{vertex}}}{\beta_0} = \frac{2 \pi e^2}{\ve_0 v_F} \left[2 f_3\left( \omega, \frac{\omega}{2v_F} \right) + f_3\left( 0, \frac{\omega}{2v_F}  \right)   + 2 f_4\left( \omega, \frac{\omega}{2v_F} \right) -  f_4\left( 0, \frac{\omega}{2v_F} \right) + 2 i  f_5
\left( \omega, \frac{\omega}{2v_F} \right)\right] \label{SMEq:deltabvertex}
\ee 
After the analytic continuation, integrals over $p$ in $f_3-f_5$ contain singularities at $p = \omega/2v_F$, which should be considered as the Cauchy principal value.

{\subsection{Total correction: Hard-cutoff regularization}}

Analogously to the case of the Hubbard interaction, the contributions to the interaction correction from both diagrams in Fig.~\ref{SMFig:bareresponse} are the same. Then, collecting together the self-energy and vertex corrections, Eqs.~(\ref{SMEq:deltabSE}) and~(\ref{SMEq:deltabvertex}), we find for the overall correction

\be 
\delta \beta = \delta \beta_{\text{self-energy}} + \delta \beta_{\text{vertex}} = \beta_0 \frac{e^2}{\pi \ve_0 v_F} F\left(  \frac{v_F q_0}{\omega} , \frac{|\mu_1|}{\omega}   \right). \label{SMEq:deltabetahardcutoff}
\ee
Function $F$ reads as 

\begin{multline}
F \equiv 2 \pi^2 \left[ - 3 f_2 \left( \frac{\omega}{2v_F} \right) -  \frac{\omega}{2v_F}  f'_2 \left( \frac{\omega}{2v_F} \right) + 2 f_3\left( \omega, \frac{\omega}{2v_F} \right) + f_3\left( 0, \frac{\omega}{2v_F}  \right) + \right. \\ \left.  + 2 f_4\left( \omega, \frac{\omega}{2v_F} \right) -  f_4\left( 0, \frac{\omega}{2v_F} \right) + 2 i  f_5
\left( \omega, \frac{\omega}{2v_F} \right)\right]. \label{SMEq:Fdef}
\end{multline}

The explicit expression for function $F$ depends on the way we regularize the UV divergencies in the theory.  In particular, the presence of an explicit UV cutoff $\Lambda$
in the momentum integrals in Eqs.~(\ref{SMEq:f1-f2}) and (\ref{SMEq:f3-f5}) corresponds to the hard cutoff (hc) regularization scheme.  This scheme implies that we neglect the electron states with momenta $p$ exceeding $\Lambda$, i.e., add the factor $\Theta(\Lambda - p)$ to the electron's Green's function, where $\Theta(x)$ is the Heaviside step function. Within this scheme, the expression for $F(x,y)$ has form
\begin{multline}
F^{\text{hc}}(x,y) = \text{p.v.} \int_{y}^{v_F\Lambda/\omega} \frac{2p \, dp}{(4p^2 -1)} \times \\ \times \int_{-1}^{1} dt \frac{-64 p^5 t - 64 p^3 x^2 t +  4 p^2 (1 + 8 x^2 - 5 t^2) + (1 + 4 x^2) (-1 + t^2) +  32 p^4 (1 + t^2) - 4 p t (-2 - 8 x^2 + t^2)}{(1 + 4 p^2 + 4 x^2 - 4 p t)^2}. \label{SMEq:Fhardcutoff}
\end{multline}
The presence of a hard cutoff $v_F \Lambda/\omega$ in this expression allows us to integrate over $t$ first. It is straightforward to show that the resulting expression, though cumbersome, is free of the UV divergence, hence, one can simply set $\Lambda \to \infty$ eventually (but only after integrating over $t$). The behavior of $F^{\text{hc}}(x,y)$ at different $y$ (different chemical potential) can be extracted from Fig.~\ref{Fig:Ffunction} and Eq.~(\ref{Eq:FscFdrFhc}) of the main text, as well as Fig.~\ref{SMFig:F(omega)}. Note that function $F$ in Figs.~\ref{Fig:Ffunction} and~\ref{SMFig:F(omega)} is obtained within the soft-cutoff regularization, see next section.

It is possible to derive simple analytical expressions for $F^{\text{hc}}$ in some limiting cases. In particular, we find

\be  
F^{\text{hc}}(0, 0) = 2, \qquad F^{\text{hc}}(x\gg 1, y<1/2) \approx \int_0^{\infty} \frac{2 p x^2 dp}{(p^2 + x^2)^2} = 1. \label{SMEq:Fhardlimits}
\ee 

While $F(x,y)$ calculated here does not explicitly depend on the UV cutoff $\Lambda$, we implied the presence of a hard cutoff, which allowed us to integrate over the angle variable $t = \cos \theta$ before integrating over $p$. This approach is somewhat rude and can be shown to violate the Ward-Takahashi identity. We find that the soft-cutoff and the dimensional regularization schemes, while giving qualitatively similar result, lead to a different quantitative answer. 

\vspace{0.5cm}

\subsection{Total correction: Soft-cutoff regularization}

 The soft-cutoff (sc) regularization scheme implies that instead of introducing a hard cutoff $\Lambda$ in the electron's spectrum (i.e., neglecting electron's states with momenta exceeding $\Lambda$) we modify the interaction potential~(\ref{SMEq:Yukawapotential}), such that it cuts off the high-momenta modes:

\be  
V(\bq) \to \frac{4\pi e^2}{\ve_0(q^2 + q_0^2)} \exp \left( - \frac{q^2}{\Lambda^2} \right), \qquad \Lambda \to \infty. \label{SMEq:softcutoffV}
\ee

The expression for the interaction correction looks very similar to the one obtained above with the hard cutoff, Eq.~(\ref{SMEq:Fdef}), with the only difference that $f_i(i\omega, k)$ from Eqs.~(\ref{SMEq:f1-f2}) and~(\ref{SMEq:f3-f5}), $i=2-5$, should now be replaced now with $f_i^{\text{sc}}(i\omega, k)$ according to 

\be 
f_i(i\omega, k) = \int_{|\mu_1|/v_F}^{\Lambda} dp \int_{-1}^{1} dt \ldots \qquad \longrightarrow \qquad f_i^{\text{sc}}(i\omega, k) = \int_{|\mu_1|/v_F}^{\infty} dp \int_{-1}^{1} dt \exp\left( - \frac{p^2 + k^2 - 2 p k t}{\Lambda^2}  \right)\times\ldots \label{SMEq:f-g}
\ee 
The contribution to $\delta \beta$ from small momenta $p\lesssim \max\{ q_0, \omega/v_F  \}$ does not depend on the UV regularization procedure and is given by Eqs.~(\ref{SMEq:deltabetahardcutoff})-(\ref{SMEq:Fhardcutoff}).  The soft cutoff, however, additionally accounts for the states with high momenta $p\sim \Lambda$, which cannot be neglected. To calculate the corresponding contribution, we use the expression analogous to Eqs.~(\ref{SMEq:deltabSE}) and~(\ref{SMEq:deltabvertex}), but with all $f_i$ being replaced by $f_i^{\text{sc}}$. Expanding then at $p\sim \Lambda \to \infty$ and keeping the leading order in $p$ terms (which is equivalent to subtracting the regular low-momentum contribution), we find

\begin{align}
F^{\text{sc}}(x,y) - F^{\text{hc}}(x,y) & =  \int\limits_{\sim \max\{\omega/v_F, q_0 \}}^{\infty} dp \int\limits_{-1}^{1} dt \left(  2t - \frac{1-3t^2}{p} - \frac{3t^2 p}{\Lambda^2}\right)\exp\left( - \frac{p^2}{\Lambda^2} \right)\approx \nonumber \\ &\approx - \int_0^{\infty}dp \int_{-1}^1 dt \frac{3 t^2 p}{\Lambda^2}\exp\left( - \frac{p^2}{\Lambda^2} \right)= -1. \label{SMEq:Fsc-Fhc}
\end{align}
Consequently, the interaction correction in the case of soft cutoff is given by

\begin{figure}
  \centering
  \caption{  The dependence of function $F$, Eq.~(\ref{Eq:deletabetaYukawa}), on $\omega$ at  fixed $\mu_1$ and different values of $q_0$ for the cases of the soft-cutoff and the dimensional regularizations. At $\omega\gg|\mu_1|,$ all curves approach $F=1$. }
  \includegraphics[width=.6\columnwidth]{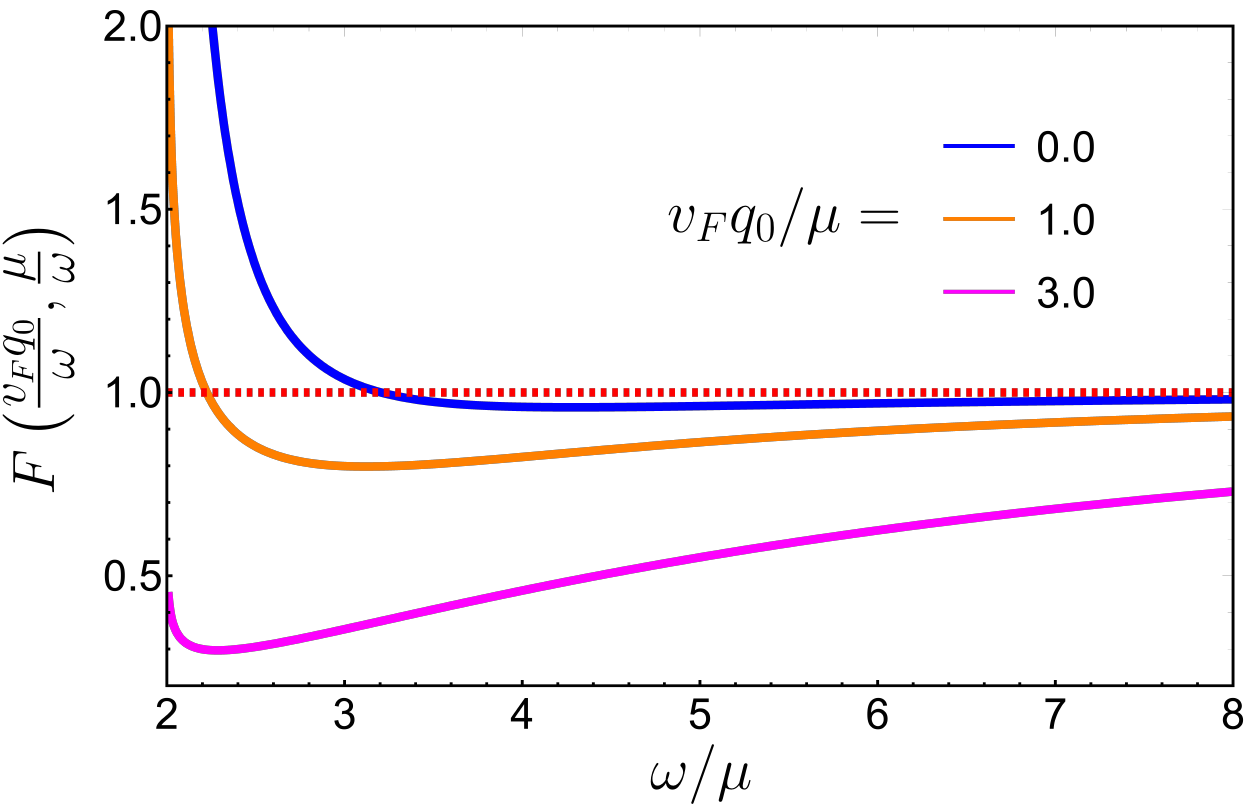}
  \label{SMFig:F(omega)}
\end{figure}

\be 
\delta \beta^{\text{sc}} =\beta_0 \frac{e^2}{\pi \ve_0 v_F} F^{\text{sc}}\left(  \frac{v_F q_0}{\omega} , \frac{|\mu_1|}{\omega}   \right) = \beta_0 \frac{e^2}{\pi \ve_0 v_F} \left[ F^{\text{hc}}\left(  \frac{v_F q_0}{\omega} , \frac{|\mu_1|}{\omega}   \right) - 1 \right]. \label{SMEq:deltabetasoftcutoff}
\ee
Hence, the soft-cutoff regularization scheme gives an anomalous high-momentum contribution to $\delta \beta$ compared to the hard cutoff, Eq.~(\ref{SMEq:deltabetahardcutoff}), which, however, does not depend to $q_0$, laser frequency $\omega$, or chemical potential $\mu_1$.

The dependence of $F^{\text{sc}}$   on frequency at fixed $\mu_1$ and different (also fixed) values of $q_0$ is shown in Fig.~\ref{SMFig:F(omega)}. All curves have logarithmic divergence when $\omega$ approaches $2|\mu_1|$, indicating that the next-order corrections must be taken into account when $\omega$ is exponentially close to $2|\mu_1|$.

It can be easily shown that the exact form of the soft cutoff in Eq.~(\ref{SMEq:softcutoffV}) does not play any role: it can be $\exp(-q^2/\Lambda^2)$, $\exp(-q/\Lambda)$, $\Theta(\Lambda - q),$ or any other function of $q$ that decays fast enough (at least power-law) to zero at $q\gtrsim \Lambda$. All of them obey the Ward-Takahashi identity (charge conservation) and lead to the same answer~\cite{SheehySchmalian2009}.

\subsection{Total correction: Dimensional regularization}

Finally, we calculate the CPGE correction $\delta \beta$ using the dimensional regularization (dr). The idea of the method is to perform the calculation in $d=3-\ve$ dimensions and take the limit $\ve \to 0$ at the end~\cite{PeskinSchroederbook}. The self-energy and the vertex corrections are again given by the expressions analogous to Eqs.~(\ref{SMEq:SigmaYukawa}) and~(\ref{SMEq:deltasigma}). However, because of the relation $\sigma^i \sigma^{\alpha} \sigma_i = -(1-\ve) \sigma^{\alpha}$, functions $f_2-f_5$ (in particular, $f_3$ and $f_4$) need to be replaced with $f_2^{\text{dr}}-f_5^{\text{dr}}$:

\begingroup
\allowdisplaybreaks
\begin{align}
f_2^{\text{dr}}(k) & = \frac1k \sum_{p} \frac{\hat \bp \cdot \hat \bk}{(\bk - \bp)^2+q_0^2}, \nonumber \\
f_3^{\text{dr}}(i\omega, k) &=  \sum_{p} \frac{2v_F^2 p }{\left[(\bk - \bp)^2+q_0^2\right] \left[ 4 v_F^2 p^2 + \omega^2 \right]} \cdot \frac{(2d-3) - (d-2)(\hat \bp \cdot \hat \bk)^2}{d-1}  \nonumber \\  f_4^{\text{dr}}(i\omega, k) &= \sum_{p} \frac{2v_F^2 p }{\left[(\bk - \bp)^2+q_0^2\right] \left[ 4 v_F^2 p^2 + \omega^2 \right]} \cdot \frac{d(\hat \bp \cdot \hat \bk)^2 - 1}{d-1} \nonumber \\  f_5^{\text{dr}}(i\omega, k) &= \sum_{p} \frac{ 2 v_F \omega (\hat \bp \cdot \hat \bk)}{\left[(\bk - \bp)^2+q_0^2\right] \left[ 4 v_F^2 p^2 + \omega^2 \right]}. 
\end{align}
\endgroup
Summation over $\bp$ also must be performed in $d$ dimensions. To do that, we introduce Feynman parameters $(\eta,\xi)$ and use the usual for the dimensional regularization $d$-dimensional (Euclidean) integrals 

\begin{align}
\int \frac{d^d p}{(2\pi)^d} \frac1{(p^2 + \Delta)^n} = \frac1{(4\pi)^{d/2}} \cdot \frac{\Gamma\left( n - \frac{d}2  \right)}{\Gamma(n)} \cdot\frac1{\Delta^{n - (d/2)}}, \nonumber \\  \int \frac{d^d p}{(2\pi)^d} \frac{p^2}{(p^2 + \Delta)^n} = \frac1{(4\pi)^{d/2}} \cdot \frac{d}2 \cdot \frac{\Gamma\left( n - \frac{d}2 -1  \right)}{\Gamma(n)} \cdot \frac1{\Delta^{n - (d/2)-1}},
\end{align}
where $\Gamma(x)$ is the Euler gamma function. We find in the limit $\ve \to 0$ for $\mu_1 = 0$:

\begingroup
\allowdisplaybreaks
\begin{align}
f_2^{\text{dr}}(k) & = \frac{\Gamma\left( \frac{3-d}2 \right)}{\sqrt{\pi} (4\pi)^{d/2}} \int_0^1 \frac{d\eta  \, \eta}{\sqrt{1-\eta} \, \Delta_2^{(3-d)/2}}  \approx \frac1{3 \pi^2 \ve}  + \frac1{8 \pi^2} \int_0^1 \frac{d\eta \, \eta}{\sqrt{1-\eta}} \left[ \ln \frac{4\pi}{\Delta_2} - \gamma \right], \nonumber \\ f_3^{\text{dr}}(i\omega, k) &= \frac{\Gamma\left( \frac{3-d}2 \right)}{2\sqrt{\pi} (4\pi)^{d/2}} \int_0^1 d\xi \int_0^{1-\xi} \frac{d\eta }{\sqrt{1- \eta - \xi} \, \Delta^{(5-d)/2}} \left[ (d-1) \Delta + \frac{3-d}2 \eta^2 k^2 \right] \approx   \nonumber \\ &\approx \frac1{3 \pi^2 \ve} + \frac1{16 \pi^2} \int_0^1 d\xi \int_0^{1-\xi} \frac{d\eta}{\sqrt{1-\eta-\xi}} \left[ \frac{\eta^2k^2}{\Delta}  +2\ln\frac{4\pi}{\Delta} - 2\gamma -2\right], \nonumber  \\  f_4^{\text{dr}}(i\omega, k) &= \frac{\Gamma\left( \frac{5-d}2 \right)}{2\sqrt{\pi} (4\pi)^{d/2}} \int_0^1 d\xi \int_0^{1-\xi} \frac{d\eta \,  \eta^2 k^2  }{\sqrt{1- \eta - \xi} \, \Delta^{(5-d)/2}}  \approx \frac{1}{16 \pi^2} \int_0^1 d\xi \int_0^{1-\xi} \frac{d\eta \,  \eta^2 k^2   }{\sqrt{1- \eta - \xi} \, \Delta},   \nonumber \\  f_5^{\text{dr}}(i\omega, k) &= \frac{\Gamma\left( \frac{5-d}2 \right)}{2\sqrt{\pi} (4\pi)^{d/2}} \cdot \frac{\omega k}{v_F}\int_0^1 d\xi \int_0^{1-\xi} \frac{d\eta \, \eta }{\sqrt{1- \eta - \xi} \, \Delta^{(5-d)/2}} \approx \frac{1}{16 \pi^2} \cdot \frac{\omega k}{v_F}\int_0^1 d\xi \int_0^{1-\xi} \frac{d\eta  \, \eta  }{\sqrt{1- \eta - \xi} \, \Delta}, \label{SMEq:r2-r5}
\end{align}
\endgroup
with
\begin{align}
\Delta_2 &  \equiv \eta(1-\eta)k^2 + \eta q_0^2,  \nonumber  \\ \Delta &\equiv \eta(1-\eta)k^2 + \eta q_0^2 + \xi\frac{\omega^2}{4v_F^2},
\end{align}
and $\gamma$ is the Euler's constant.

In general, correlators~(\ref{SMEq:corr0}) and~(\ref{SMEq:corr1}) should also be evaluated in $d$ dimensions (which can be done readily). However, since the $1/\ve$ poles from $f_2^{\text{dr}}$ and $f_3^{\text{dr}}$ (both of which are multiplied by 
$\langle \tr\left[ \sigma^{\alpha} P_-(\bk) \sigma^{\beta} P_+(\bk) \sigma^{\gamma} P_+(\bk) \right]\rangle_{\hat \bk}$) exactly cancel each other, while $f_4^{\text{dr}}$ and $f_5^{\text{dr}}$ do not have $1/\ve$ poles at all, it is sufficient to calculate all the correlators exactly at $d=3$. Consequently, the function $F^{\text{dr}}$  that determines the correction to the CPGE coefficient, Eq.~(\ref{SMEq:deltabetahardcutoff}), is given by the same Eq.~(\ref{SMEq:Fdef}), but with all $f_i$ now replaced with $f_i^{\text{dr}}$.

A word of caution is in order here. Strictly speaking, the correct coefficient at $f_2$ in Eq.~(\ref{SMEq:Fdef}) is $3-\ve$, not $3$. However, this extra $\ve f_2$ term (which is of the order 1) is exactly compensated by the bare CPGE coefficient, if we take the renormalization of the Fermi velocity into account~\cite{Teber2017,TeberKotikov2018}. Indeed, within the dimensional regularization in $3-\ve$ dimensions, the bare CPGE coefficient equals 

\be
\beta_0^{(3-\ve)} = \frac{i e^3}{12 \pi} v_F^{\ve} \approx \frac{i e^3}{12 \pi} (1 + \ve \ln v_F), \label{SMEq:beta0(3-epsilon)}
\ee
where we omitted the factor  $\Theta(\omega - 2|\mu_1|)$ for brevity and neglected the terms that vanish in the limit $\ve \to 0$. Na\"{i}vely, the second term in Eq.~(\ref{SMEq:beta0(3-epsilon)}), should be set to 0 in the limit $\ve \to 0$ as well. However, it is crucial to recall that $v_F$ in the above expression is the bare Fermi velocity. The renormalization procedure of the field theory, on the other hand, requires this bare $v_F$ to be expressed through the renormalized (running) Fermi velocity $\tilde v_F$, which, to the leading order in $e^2/v_F \ve_0$, are related as
\be
\tilde v_F \approx  v_F\left(1 + \frac{2\pi e^2}{\ve_0 v_F} f_2\right), \qquad \text{or} \qquad  v_F \approx  \tilde v_F\left(1 - \frac{2\pi e^2}{\ve_0 \tilde v_F} f_2\right).
\ee
Hence, re-expressing the bare CPGE coefficient in $d = 3- \ve$ dimensions through the renormalized Fermi velocity, we find

\be 
\beta_0^{(3-\ve)}  \approx \frac{i e^3}{12 \pi} \left(1 + \ve \ln \tilde v_F-\frac{2\pi e^2}{\ve_0 \tilde v_F} \ve f_2 \right) \approx \frac{i e^3}{12 \pi} \left(1 -\frac{2\pi e^2}{\ve_0  v_F} \ve f_2 \right),
\ee
where we used $\ve \ln \tilde v_F \to 0$ at $\ve \to 0$, and changed $\tilde v_F \to v_F$ in the last term (which does not make any difference to the leading order). It is straightforward to show that the additional $\ve f_2$ correction in $\beta_0^{(3-\ve)}$ exactly cancels the extra $\ve f_2$ term originating from the self-energy.  Consequently, all the coefficients in Eq.~(\ref{SMEq:Fdef}) turn out to be correct. The above argument can be made more rigorous by expressing everything through the renormalized (physical) parameters and introducing the counter-terms which absorb all the UV divergencies~\cite{PeskinSchroederbook}.

Next, it can be shown  that $F^{\text{dr}}(x,y)$ is related to its hard-cutoff counterpart, $F^{\text{hc}}(x,y)$, by a simple expression 

\be 
F^{\text{dr}}(x,y) = F^{\text{hc}}(x,y) - 1, \label{SMEq:Fdim-Fhard}
\ee
analogously to the soft-cutoff result~(\ref{SMEq:Fsc-Fhc}). The easiest way to obtain the above relation  analytically is to consider the limit $\mu_1 = 0,$ $\omega \ll q_0$, which corresponds to $y=0$ and $x\to \infty$. In this limit, one has $\Delta_2 = \Delta = \eta q_0^2$, and we find

\begin{align}
& f_2^{\text{dr}} \approx f_3^{\text{dr}} \approx \frac1{3 \pi^2 \ve} + \frac1{6 \pi^2} \left[ \ln \frac{4\pi}{q_0^2} - \gamma  - 2 \ln2 + \frac53 \right], \nonumber \\ & f_4^{\text{dr}} \approx f_5^{\text{dr}} \approx \frac{\omega}{2v_F} \cdot \left(f_2^{\text{dr}}\right)' =0,
\end{align}
leading to $F^{\text{dr}}(x\to \infty,0) = 0$, in full agreement with Eqs.~(\ref{SMEq:Fhardlimits}) and~(\ref{SMEq:Fdim-Fhard}). 

Different regularizations differently account for the high-energy states only, while the regular contribution is the same for all schemes. It implies that different functions $F$ can only differ by a constant, which can be found in any convenient limit. This means, in turn, that Eq.~(\ref{SMEq:Fdim-Fhard}) holds not only in the limit $y=0$ and $x\to \infty$ which we have considered explicitly but also for arbitrary $x$ and $y$.

We have demonstrated that the results for the interaction correction obtained within the soft-cutoff and the dimensional regularization schemes are the same, while the one with the hard cutoff is different. Though not explicitly dependent on the UV cutoff $\Lambda$, this discrepancy originates from the high-energy states with momenta $p\sim \Lambda$. Thus, it is not surprising that the hard-cutoff procedure fails to give the correct answer, since it violates the Ward-Takahashi identity and, consequently, does not accurately account for these high-energy states.

\end{widetext}

\end{document}